\documentclass[12pt,nofootinbib,aps,amsmath,amssymb,showpacs]{revtex4}

\usepackage[T1]{fontenc} \usepackage[latin1]{inputenc}
\usepackage{amsmath,color,ulem} \usepackage{graphicx}
\usepackage{epsf,bm} \usepackage{amssymb}

\usepackage{tabulary}
\newcolumntype{K}[1]{>{\centering\arraybackslash}p{#1}}
\usepackage{multirow}
\usepackage{float}
\usepackage{hyperref}
\hypersetup{
	colorlinks=true,    
	linkcolor=blue,     
	citecolor=blue,    
	urlcolor=blue        
}

\begin{document}
	
	\title{Domain coarsening in polycrystalline graphene}
	\author{Zihua Liu$^\dagger$} 
	\email{z.liu1@uu.nl}
	\author{Debabrata Panja$^\dagger$}
	\author{Gerard T. Barkema$^\dagger$}

	\affiliation{
		$^\dagger$Department of Information and Computing Sciences, Utrecht 
		University, Princetonplein 5, 3584 CC Utrecht, The Netherlands\\
	}
	
	\date{\today} 
\begin{abstract}
	Graphene is a two-dimensional carbon material which exhibits
  exceptional properties, making it highly suitable for a wide range
  of applications. Practical graphene fabrication often yields a
  polycrystalline structure with many inherent defects, which
  significantly influence its performance. In this study, we utilize a
  Monte Carlo approach based on the optimized Wooten, Winer and Weaire
  (WWW) algorithm to simulate the crystalline domain coarsening
  process of polycrystalline graphene.  Our sample configurations show
  excellent agreement with experimental data. We conduct statistical
  analyses of the bond and angle distribution, temporal evolution of
  the defect distribution, and spatial correlation of the lattice
  orientation that follows a stretched exponential
  distribution. Furthermore, we thoroughly investigate the diffusion
  behavior of defects and find that the changes in domain size follow
  a power-law distribution. We briefly discuss the possible
  connections of these results to (and differences from) domain growth
  processes in other statistical models, such as the Ising
  dynamics. We also examine the impact of buckling of polycrystalline
  graphene on the crystallization rate under substrate effects. Our
  findings may offer valuable guidance and insights for both
  theoretical investigations and experimental advancements.
\end{abstract}

\maketitle

\section{Introduction}

Graphene is a material in which carbon atoms are organized in the
structure of a honeycomb lattice.  It exhibits a wide range of
appealing properties in comparison to more conventional materials,
including exceptionally high strength and toughness
\cite{song2020tailoring,zhang2014fracture,shekhawat2016toughness},
remarkable thermal conductivity
\cite{balandin2008superior,chen2012thermal}, and outstanding
electrical conductivity \cite{chen2008mechanically}.  As a result, use
of graphene-based devices has witnessed a substantial surge in recent
years
\cite{wang2009supercapacitor,gwon2011flexible,chen2013terahertz,zhang2013high}.
Graphene can be fabricated experimentally through different methods,
such as chemical vapor deposition (CVD) \cite{zhang2013review},
epitaxial growth on silicon carbide \cite{mishra2016graphene}, and
liquid-phase exfoliation \cite{xu2018liquid}. However, graphene
produced using these techniques typically exists in a polycrystalline
form, which means that the structure consists of many crystalline
domains, each with its own lattice orientation. Neighboring domains
are separated by strings of defects, usually five-fold and seven-fold
rings.  A sample of polycrystalline graphene is depicted in
Fig. \ref{fig1}(d).

The structure of polycrystalline graphene is not stationary in
time. Changes in the bonded structure occur all the time via the
so-called bond translocations. If such a bond translocation occurs in
the middle of a crystalline region, four six-fold rings evolve into
two five- and two sevenfold rings [middle panel of
Fig. \ref{fig1}(b)], a structure known as a Stone-Wales
defect. Occasionally arising Stone-Wales defects in otherwise
crystalline graphene tend not to last, and in due time, the
crystalline structure is restored.  If a bond translocation occurs in
the immediate vicinity of a five- and sevenfold ring, the result is
that this 5-7 pair is actually displaced sideways [right panel of
Fig. \ref{fig1}(b)].  Via this mechanism, the domain walls separating
the crystalline regions, consisting of alternating strings of five-
and sevenfold rings, can actually wander.

The global effect of this wandering of the domain walls is {\it
  coarsening} or {\it domain growth}: bigger domains tend to grow at
the expense of smaller ones, because of energetic considerations, and
the density of domains decreases in time.

Here we study the domain growth process in graphene using computer
simulations.  First, in order to understand the force that drives the
coarsening process, we study the energetics of polycrystalline
graphene: in particular, we show that the total energy of the system
increases monotonically with the number of 5- and 7-fold rings in a
more or less linear fashion (Fig. \ref{fig4}).  Next, we study the
evolution in time of the defect density, spatial correlation of the
lattice orientation and the average domain size.  We find that the
defect density scales as $t^{-1/3}$ in flat polycrystalline graphene,
the spatial correlation of the lattice orientation is well fitted by a
stretched exponential function, and the average size of the domains
grows like $t^{1/6}$. We discuss similarities to the domain growth
process (so-called Ostwald ripening) in the Ising model. We also
investigate the influence of buckling on the coarsening process and
find that buckling of polycrystalline graphene slows it down. This
implies that graphene samples with better crystallinity are best
produced if the graphene is kept as flat as possible by a substrate.

This paper is organized as follows. First, in Sec. \ref{sec2}, we
describe our model for graphene, including its dynamics. In
Sec. \ref{newsec3}, we validate our model, and present that the
structures resulting from simulations are in good agreement with
experimental data. Next in Sec. \ref{sec3}, we present a statistical
analysis of the spread in bond lengths and bond angles, structural
disorder and defect density, as a function of time.  We also present
an extensive study of the lattice orientations, both in its spatial
distribution and its dynamics.  In Sec. \ref{sec4}, we analyses the
diffusive behavior of defects and the separation of crystal phases.
In section \ref{sec5}, we discuss the influence of binding to the
substrate for buckled polycrystalline graphene. We conclude the paper
with a summary in Sec. \ref{sec6}.

\section{Model }\label{sec2}

 \begin{figure*}
	\includegraphics[width=0.9\linewidth]{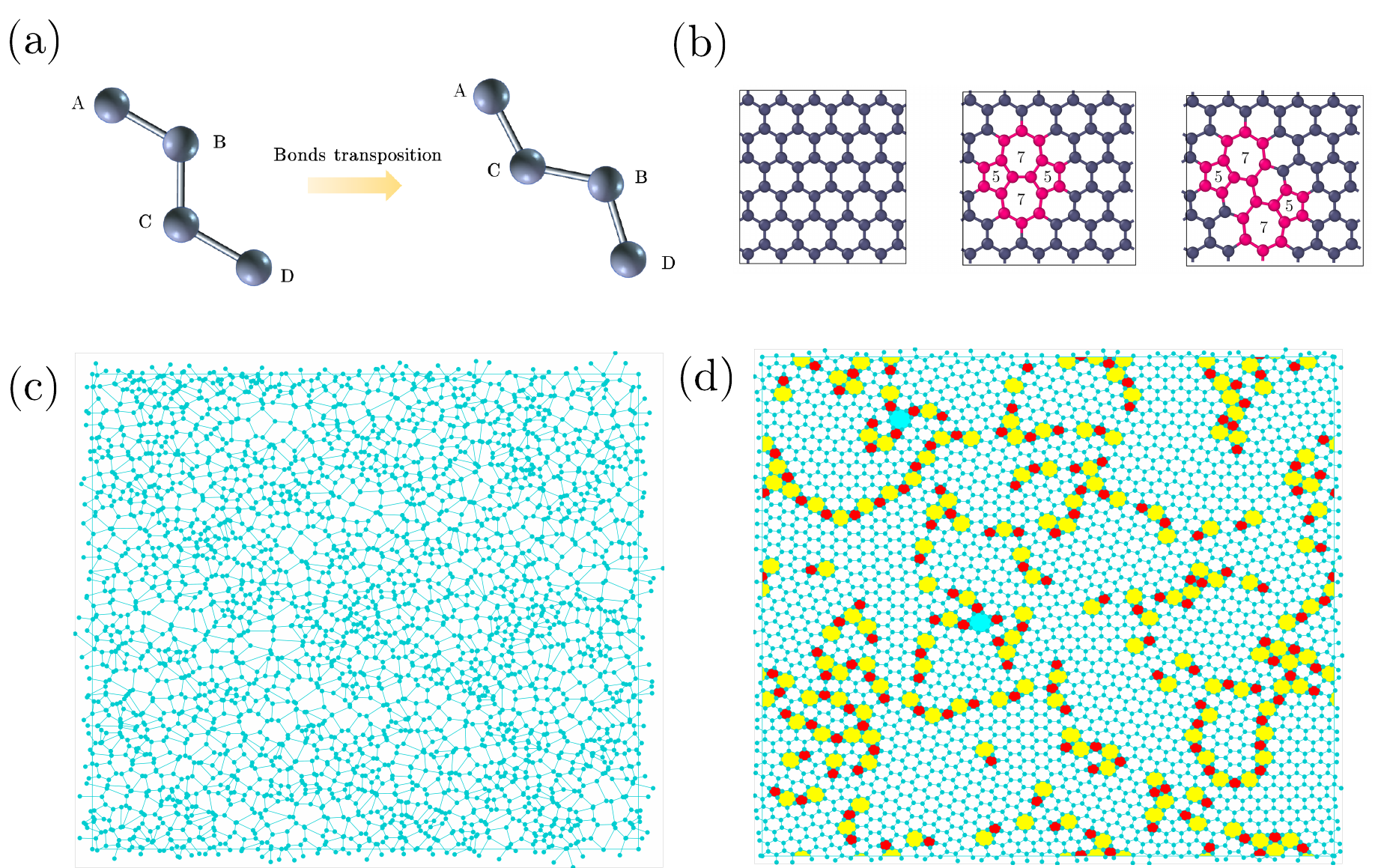}
\caption{(a) Elementary move in the structural evolution of
          polycrystalline graphene, also known as a bond
          transposition. In a string of four carbon atoms A-B-C-D, the
          bonds A-B and C-D are replaced by bonds A-C and B-D, leaving
          the central bond B-C untouched. (b) Left panel: if a bond
          transposition occurs in crystalline graphene, it results in
          two oppositely oriented pairs of 5-7 rings. Right panel: if
          a bond transposition occurs in the immediate vicinity of a
          5-7 pair, it is effectively displaced sideways. (c)
          Visualization of an initial sample, created from a Voronoi
          network as described in the text. Note that the network is
          disordered and homogeneous, with at most tiny crystalline
          regions. (d) Same network after structural relaxation with
          $9\times 10^4$ proposed bond transitions, when crystalline
          regions have appeared. In this figure, 5-, 7- and 8-fold
          rings are marked in different colors. 
          \label{fig1}}
\end{figure*} 

The overwhelming majority of carbon atoms in graphene are covalently
bonded to three neighboring atoms; undercoordinated carbon atoms do
exist, but at a density which is so low, that it can be safely
neglected. In this work, we use the recently developed semiempirical
energy potential which has the following form \cite{jain2015strong}:

\begin{align}
	\label{eq1}
	E=\frac{3}{16}\frac{\alpha}{d^2}\sum_{i,j}^{}{\left( r_{ij}^{2}-d^2 \right) ^2+\frac{3}{8}}\beta d^2\sum_{j,i,k}^{}{\left( \theta _{jik}-\frac{2\pi}{3} \right) ^2+\gamma \sum_{i,jkl}^{}{r^2_{i,jkl}}}.
\end{align}

Here, $r_{ij}$ is the distance between two bonded atoms $i$ and $j$,
$\theta_{jik}$ is the angle between the two bonds connecting atom $i$
to atoms $j$ and $k$ and $r_{i,jkl}$ is the out-of-plane distance from
atom $i$ to the plane through the three atoms $j$, $k$ and $l$.  The
parameter $\alpha$ is chosen as 26.060eV/\AA$^2$ to control bond
stretching and is fitted to the bulk modulus.  The parameter
$\beta=5.511$eV/\AA$^2$ governs bond shearing and is fitted to the
shear modulus.  The parameter $\gamma=0.517$eV/\AA$^2$ describes the
stability of the graphene sheet against buckling; note that this third
out-of-plane term is zero in perfectly flat graphene (2D simulations).
$d=1.420$\AA \space is the ideal bond length for pure graphene.  All
these parameters are obtained by fitting to density functional theory
(DFT) calculations.  Note that the elastic potential strictly requires
the bond list where each atom is bonded to exactly three atoms; the
number of bonds equals therefore $3N/2$, in which $N$ is the number of
atoms.  This potential enables one to efficiently estimate the
energies of the relatively stable configurations encountered in our
simulations of graphene coarsening.

Simulations of covalently-bonded materials are typically slow and
computationally expensive; their high stability causes the relevant
experimental time scales to be well beyond those accessible by
standard molecular dynamics simulations.  Here we employ a relatively
simple and accurate model for dynamics of polycrystalline graphene,
which was initially applied for generating silicon samples with
realistic structures.  The model constructs atomic configurations
generated by the evolution of a continuous random network (CRN) via
bond transpositions, which is a well-established and widely used
method to generate realistic atomic configurations of carbon/silicon
materials.  More specifically, we use the algorithm introduced in
Ref. \cite{barkema2000high}, an improved version of the original
method of Wooten, Winer and Weaire \cite{wooten1985computer}.  The
improved bond transposition procedure consists of the
following sequential steps: (1) constructing a comprehensive list of
bonds in the current sample configuration; (2) randomly selecting four
connected atoms (ABCD); (3) breaking the bonds between AB and CD and
forming new bonds between AC and BD, as shown in
Fig. \ref{fig1}(a); (4) performing global energy minimization and
comparing the energy $E_a$ after the bond switch with a predefined
energy threshold, defined as:

\begin{align}
	E_t=E_b-k_{\text B}T\ln \left( s \right),
\end{align}
where $k_{\text B}$ is Boltzmann constant, $T$ is temperature, $E_b$
is the energy before the bond transposition and $s$ is a uniform
random number between 0 and 1.  If the energy $E_a$ after the bond
transposition is less then $E_t$, the proposed change is accepted;
otherwise, it is rejected, and the atoms and bond list are restored.

To accelerate the evolution program, we first relax only the atoms in
the near vicinity of the bond transposition, bringing the total energy
down to $E_l$.  We then estimate the energy $E_a$ after global
relaxation (without performing the global minimization), employing a
local energy criterion in term of the linear relationship between
local minimum energy and the total remaining forces
$\left| F \right|^2$:
\begin{align}
	E_a\approx E_l-c_f\left| F \right|^2
\end{align}
Here, $c_f$ is a linear factor obtained from simulations. In our
recent work \cite{d2019discontinuous}, we found that the
performance of the local decision depends on the set of atoms allowed
to move during the local relaxation; and for this, a shell of atoms
was selected with the shortest-path distance $l$ from the atoms
involved in the bond transpositions.  In the simulations discussed in
the paper, $l$ and $c_f$ are chosen as 3 and $6\times10^{-3}$
$s^2u^{-1}$ to achieve the best performance, respectively.  Note that
the quantity $c_f$ is expressed in units of seconds squared over the
atomic mass unit.

The minimization approach exerted in our simulations is the so-called
fast inertial relaxation engine (FIRE) algorithm, in which parameters
corresponding to Ref. \cite{guenole2020assessment} are set as:
$N_{min}=5$, $f_{inc}=1.1$, $f_{dec}=0.5$, $\alpha_{start}=0.1$ and
$f_{\alpha}=0.99$.  Other custom parameters here are set to be
$\Delta t_{MD}=0.03$ and $\Delta t_{max} \sim 10\Delta t_{MD}$.  The
velocity Verlet method is chosen for the integration in time.

Fig. \ref{fig1}(c) to (d) demonstrate the evolution process of
polycrystalline graphene samples.  Fig. \ref{fig1}(c) presents a
Voronoi diagram with a random structure, providing an initial
disordered state. Note that this initial state merely provides a
homogeneous disordered network without orientational bias, and does
not reflect any practical physical process.  The construction of the
Voronoi diagram involves several steps: (1) randomly choose a set of
points within a simulation box; (2) for each seed point, determine its
region, i.e. the set of points which are nearer to it, than to another
seed point. (3) construct the boundaries of the Voronoi cells, which
are formed by the perpendicular bisectors of the lines connecting
neighboring seed points; (4) these boundaries are considered the
covalent bonds, and the positions where three of these meet are
considered as ``atom'.  Each ``atom'' within the Voronoi diagram is
strictly limited to having three neighbors, and periodic boundary
conditions (PBC) are applied to ensure a constant number of atoms
($N$) and bonds (3$N$/2) within the simulation box.
Fig. \ref{fig1}(d) shows the evolution of a polycrystalline graphene
structure with a defined defect density achieved by implementing
$9 \times 10^4$ proposed bond transpositions.  The nanocrystalline
domains with distinct crystal orientations are separated by domain
walls consisting mainly of 5- and 7-fold rings.  Further, individual
defect islands emerge within the crystalline domains.

\section{Model validation\label{newsec3}}

The model was first introduced in Ref. \cite{jain2015strong}, which is
based on Kirkwood's potential \cite{kirkwood1939skeletal}. This
potential has been used, for instance, for studying the structural
dynamics of single-layer polycrystalline
graphene\cite{liu2022structural}, for studying the long-range
relaxation of structural defects \cite{jain2015strong}, for probing
crystallinity of graphene samples via their vibrational spectrum
\cite{jain2015probing}, for the study of twisted and buckled bilayer
graphene \cite{jain2016structure} as well as of the shape of a
graphene nanobubble \cite{jain2017probing} and the discontinuous
evolution of the structure of stretching polycrystalline graphene
\cite{d2019discontinuous}.

Crystalline graphene is a two-dimensional material, but as soon as the
structure has defects --- in particular if it is polycrystalline ---
the carbon atoms tend to relief stress by {\it buckling},
i.e. displace with respect to each other in the out-of-plane
direction. For free-floating graphene in vacuum, the buckling can have
an amplitude of many angstroms, while for graphene attached to a
substrate, the amplitude of the buckling away from the substrate is
suppressed significantly.  In the first part of our simulations, we
confine the graphene to a two-dimensional plane without any buckling;
further on, we relax the constraint to the plane and allow for
buckling.

\begin{figure*}
\centering \includegraphics[width=12cm]{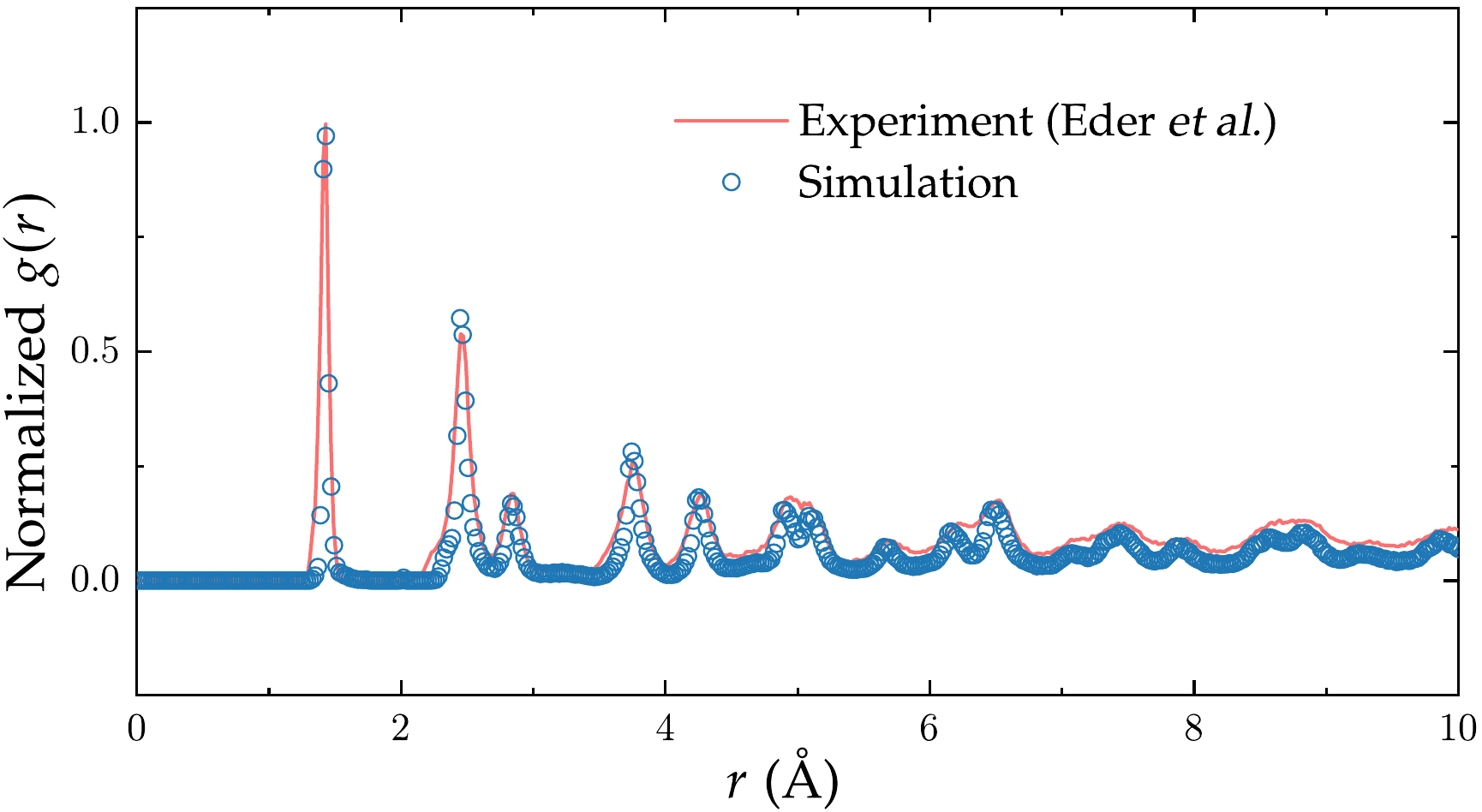}
\caption{Comparison of the normalized radial distribution function
  $g(r)$ of our generated sample and experiment, at comparable defect
  density. The two curves match very well, up to about the first ten
  peaks.\label{fig2}}
\end{figure*}

In order to ensure the validity of the obtained samples, we
employ the radial distribution function (RDF) as defined in
Eq. (\ref{rdf}), which characterizes the average spatial distribution
of particles in a system. The RDF is defined as
\begin{align}
	g\left( r \right) =\frac{\underset{\Delta r\rightarrow 0}{\lim}\frac{N_r}{\pi \left( r+\Delta r \right) ^2-\pi r^2}}{\rho},
	\label{rdf}
\end{align}
where $r$ is the radial distance from reference particle, $\rho$ is
the average atoms density and $N_r$ is the number of atoms between $r$
and $r +\Delta r$. Starting from the initial configuration
[Fig. \ref{fig1}(c)] we let the sample evolve in time. We then compare
in Fig. \ref{fig2}, the normalized radial distribution function of the
samples [Eq. (\ref{rdf})] on the two-dimensional plane when the defect
density reaches the same ($\sim$20\%) as those from
%
the experiment of Eder \textit{et al.} \cite{eder2014journey}.
(The defect density is defined as the ratio of non-hexagonal rings to
the total number of rings.) The comparison reveals an excellent
simulation-experiment agreement. Note also that the simulated samples
we used have similar long-range disorder as the ones observed in real
polycrystalline graphene.

\section{Results } \label{sec3}

\subsection{Domain growth in flat polycrystalline graphene\label{sec3a}}

\begin{figure*}[!htpb]
    \centering
    \includegraphics[width=12cm]{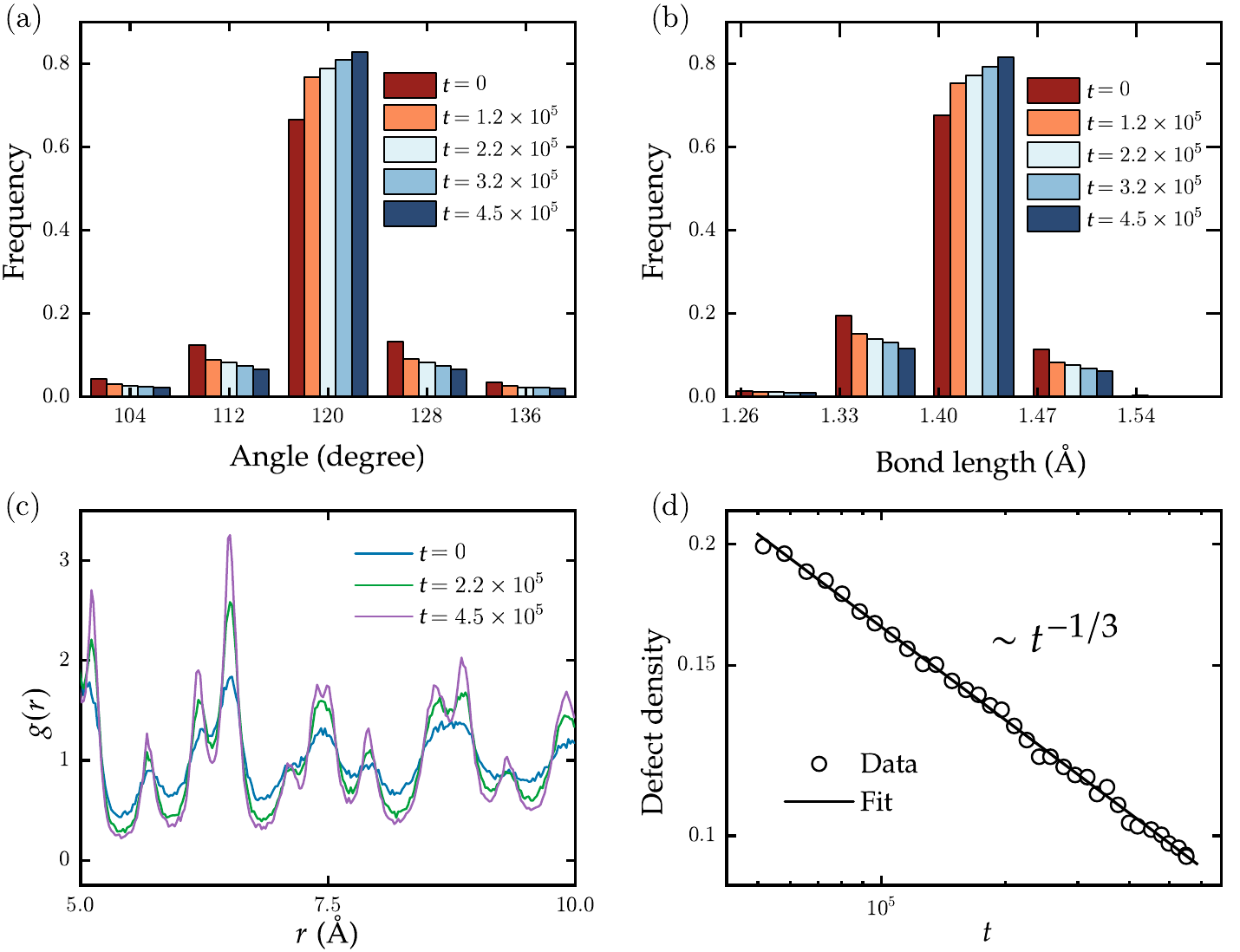}
  \caption{Time evolution of the distribution of (a) the bond
angles and (b) the bond lengths in planar samples of graphene. With
increasing simulation times, both distributions become narrower.  (c)
Time evolution of the radial distribution function. With increasing
simulation time, the peaks at longer distance become increasingly
pronounced.  (d) Density of defects (5- and 7-rings) as a function of
simulation time. The decay can be well fitted by a power-law decay
~$t^{-1/3}$ (solid line).
\label{fig3}}
\end{figure*}

For studying domain growth of realistic polycrystalline graphene
samples, at $t=0$ we start with one consisting of 9800 atoms and a
defect density of $\sim20$\%. We then evolve it for $4.5 \times 10^5$
Monte Carlo steps (MCS) under weak pressure and quench to 3000K, a
temperature significantly below the melting temperature of
polycrystalline graphene. To improve our statistics,we repeated the
evolution process 50 times using different random number seeds. The
simulations were performed on an Intel i7-9700 CPU, with an average
runtime of approximately 0.02s per MCS.  Figures \ref{fig3}(a) and
\ref{fig3}(b) display the distributions of bond angles and bond
lengths for different times, respectively. Note here that in flat
polycrystalline graphene, the third term in the potential function
[Eq. (\ref{eq1})] related to dihedral angles can be neglected due to
the absence of out-of-plane forces. Consequently, the bond angles
gradually approach the ideal value of 120$^\circ$, while the bond
lengths tend to converge to 1.42\AA.  Figure \ref{fig3}(c) illustrates
the time-dependent changes of the RDF in the range of 5\AA \space to
10\AA.  With increasing time, distinct peaks of the RDF appear at
multiple positions, indicating the gradual appearance of longer-ranged
order and an increase in the domain area. Figure \ref{fig3}(d)
displays the power-law behavior of the defect density as a function of
time, with an exponent of $-0.330\pm0.002$.  Based on this result, we
speculate that the exponent for the defect density under ideal
conditions is $-1/3$.

\begin{figure*}[!htpb]
    \centering \includegraphics[width=8cm]{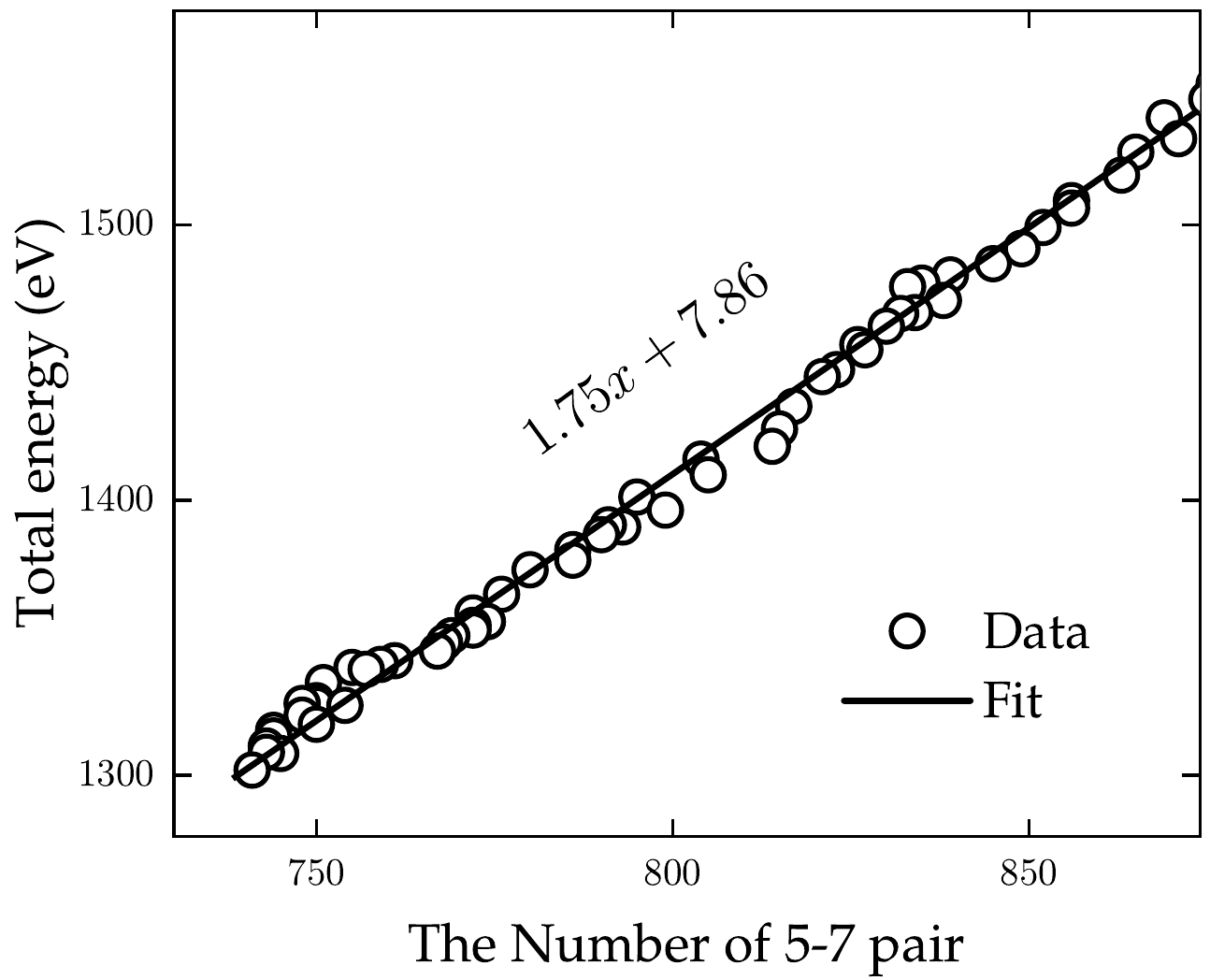}
  \caption{Total energy as a function of the number of defects (pairs
    of 5- and 7-rings) in planar graphene. The data can be well fitted
    with a linear relation: $E=1.75x+7.86$. As a reference, a single
    SW defect consists of two such pairs and would thus correspond to
    a defect formation energy of 3.5 eV.\label{fig4}}
\end{figure*}

Figure \ref{fig4} shows the linear relationship between the total
energy and the number of 5-7 pairs; it can be linearly fitted by
$f(x)=1.75x+7.86$.  As a reference, this corresponds to e formation
energy of a single Stone Wales (SW) defect in the flat polycrystalline
graphene by nearly 3.5eV \cite{jain2015strong}, as each SW defect
consists of two 5-7 pairs.

During the growth process of polycrystalline graphene, it is common to
observe the formation of domains with different lattice orientations.
The lattice orientation of these domains is complex and influenced by
various factors, such as the motion of individual defects, the
alignment of domain boundaries, and external pressure.  These factors
can exert torques to the domains, leading to a certain degree of
lattice rotation within the domains. For graphene, the range of the
lattice orientation is $-30^\circ$ to $30^\circ$, with positive values
indicating orientations corresponding to rotations around the $z$-axis
in the positive direction and negative values indicating orientations
pointing towards the negative direction of rotation around the
$z$-axis.  We applied polyhedral template matching (PTM) to identify
the lattice orientation of atoms in polycrystalline graphene
\cite{larsen2016robust}.  This method enabled one to classify
structures according to the topology of the local atomic environment,
without any ambiguity in the classification, and with greater
reliability than, e.g., common neighbor analysis in the presence of
thermal fluctuations. It is important to note the custom parameter
root-mean squared deviation (RMSD), a higher RMSD cutoff will lead to
more identifications (and fewer defect atoms), though possibly at the
expense of false positives. A lower RMSD cutoff will result in fewer
structural identifications (and more defect atoms and greater
sensitivity to perturbations of the lattice), though possibly at the
expense of false negatives. The RMSD has been set to $0.1$ in our
simulations to achieve optimal identification results.  With this
setting, the hexagonal lattice structure and defects can be identified
relatively accurately. However, the identification performance for
defects is not as robust as the ring identification algorithm used in
the previous text, which can identify non-hexagonal ring defects with
100\% accuracy.

\begin{figure}[H]
	\centering
	\includegraphics[width=13cm]{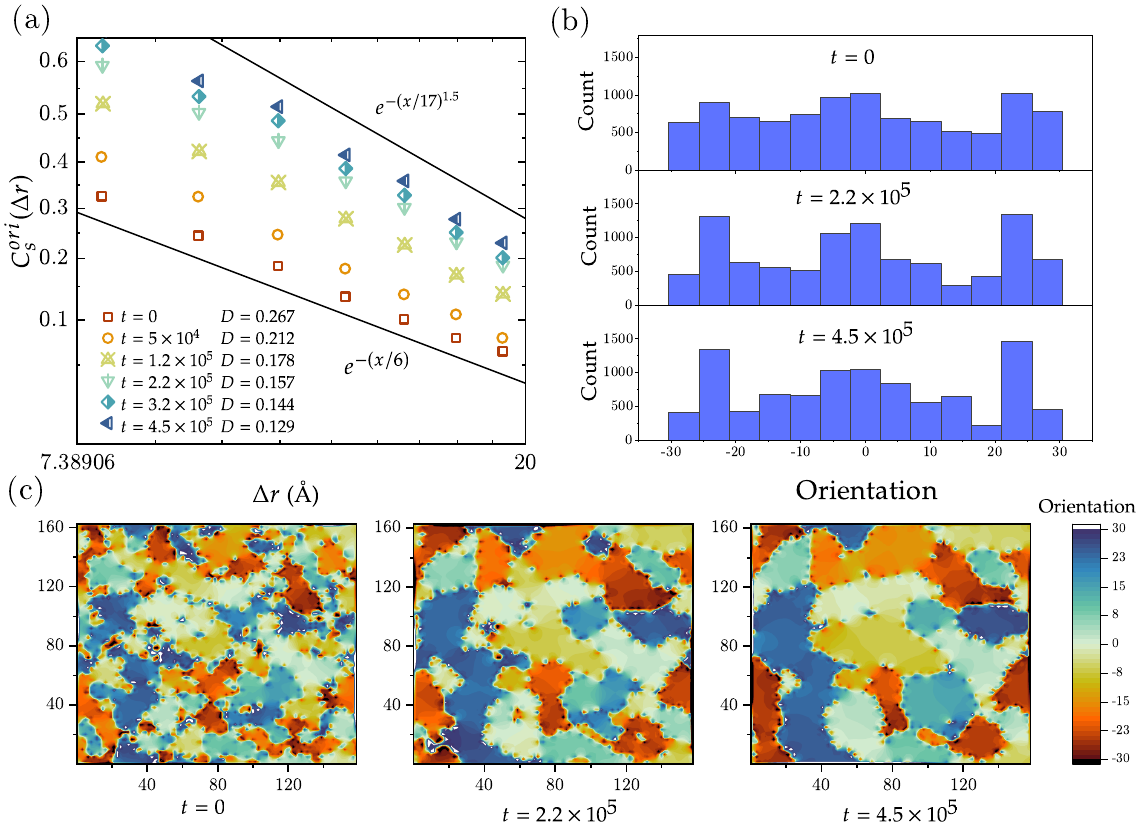}
	\caption{Analysis of the orientation of the hexagons in the lattice
		structure. (a) Normalized correlation function $C_s^{\text{ori}}$ of 
		the 
		orientations as defined in Eq. (\ref{co}), as a function of
		distance $r$, for various times. The data show a linear trend if
		$(-\ln(C_s^{\text{ori}}))$ is plotted as a function of distance $r$ in a
		double-logarithmic plot, indicating that the correlation function
		$C_s^{\text{ori}}$ decays as a stretched-exponential. (b) Histogram of
		the hexagon orientations at different times. While the crystalline
		regions grow in time, these histograms become increasingly
		rugged. (c) Evolution of the maps of the hexagon
		orientations. Some regions grow (while conserving their
		orientation), at the expense of other regions that shrink and
		sometimes disappear. 
		\label{fig5}}
\end{figure}

In order to investigate the spatial distribution of the lattice
orientation, we define the normalized spatial correlation of the
lattice orientation $C_s^{\text{ori}}$ below,
\begin{align}
  C_{s}^{\text{ori}}\left( \Delta r \right) =\frac{\left< o_i\times o_j \right>}{\left< o_{i}^{2} \right>},
  \label{co} 
\end{align}
where $o_i$ and $o_j$ are the orientation of atoms $i$ and $j$,
respectively, $\vec{r}_i$ and $\vec{r}_j$ are the corresponding
position, and with the fixed distance
$\Delta r=\left| \vec{r}_i-\vec{r}_j\right|$.  Figure \ref{fig5}(a)
shows the variation of $C_s^{\text{ori}}$ as a function of $\Delta r$
at different times $t$ and corresponding defect density $D$.  The
vertical axis shows $\ln[-\ln(C_s^{\text{ori}})]$, while the horizontal axis is
logarithmically scaled.  The data exhibits a straight decay pattern in
the figure, suggesting a trend that follows stretched exponential
decay with a form like $ C_s^{\text{ori}} \sim e^{-(\Delta r/b)^c}$.  Due to
the limitations of sample size and the effects of periodic boundary
conditions, there is a significant amount of noise present on spatial
scales larger than 20 \AA.  As a result, it becomes difficult to
present the spatial correlation lattice orientation at larger
scales. As shown in Fig. \ref{fig5}(a), the reference lines
indicate that an anomalous exponent $c$ is observed in the range from
$1$ to $1.5$.

The histogram distribution plots in Fig. \ref{fig5}(b) illustrate the
quantitative analysis of lattice orientations at three different
times. Evidently, that there is a symmetry around zero orientation,
indicating that the polycrystalline graphene can be regarded as a
binary mixture composed of two types of regions: those with
orientations greater than zero degrees and those with orientations
less than zero degrees, in equal proportions. The average size of the
binary mixture corresponds to the intersection between the correlation
curve and the $x$-axis in Fig. \ref{fig5}(a).  In the histogram, three
prominent peaks are observed in the intervals
($-30^\circ$)-($-20^\circ$), ($-10^\circ$)-($10^\circ$), and
($20^\circ$)-($30^\circ$), suggesting a higher concentration of atomic
orientations within these ranges.  Further, we observe that as time
progresses from $t=0$ to $t=4.5 \times 10^5$ in terms of MCS, the
intensity of these peaks increases, which is also in line with the
lattice orientation distribution map shown in Fig. \ref{fig5}(c).
There is a trend suggesting that smaller regions with the same
orientation have a higher tendency to merge into larger regions, and
regions with similar orientations are more prone to fusion.

\subsection{Dynamics of crystal phases\label{sec4}}

In the samples of polycrystalline graphene, crystal phases can be
identified, each consisting of carbon atoms organized in a honeycomb
lattice structure, with an orientation that differs from one domain to
another.  At the boundaries between domains, the three-fold
coordination of the bond structure is preserved, but the honeycomb
structure is discontinued by the presence of strings of 5- and 7-fold
rings.

Identifying different crystal phases and their orientations can be
challenging.  In our simulations, we employ a method called graph
clustering to identify the phases and their orientations.  This
approach is sensitive enough to detect sub-phases with subtle
differences in domain orientations.  The local structural environment
and orientation of each atom is determined using the PTM algorithm,
then graph edge weights are initialized as $\exp(-d^2/3)$, where $d$ is
the misorientation in degrees between two neighboring atoms. Domains
are built up by contracting edges using the Node Pair Sampling method
of Bonald $et\ al.$ \cite{bonald2018hierarchical}.  In our
simulations, two important parameters, the merge threshold and the
minimum grain size are set to 11 and 10, respectively, to achieve the
best performance.
\begin{figure}[H]
    \centering
    \includegraphics[width=13cm]{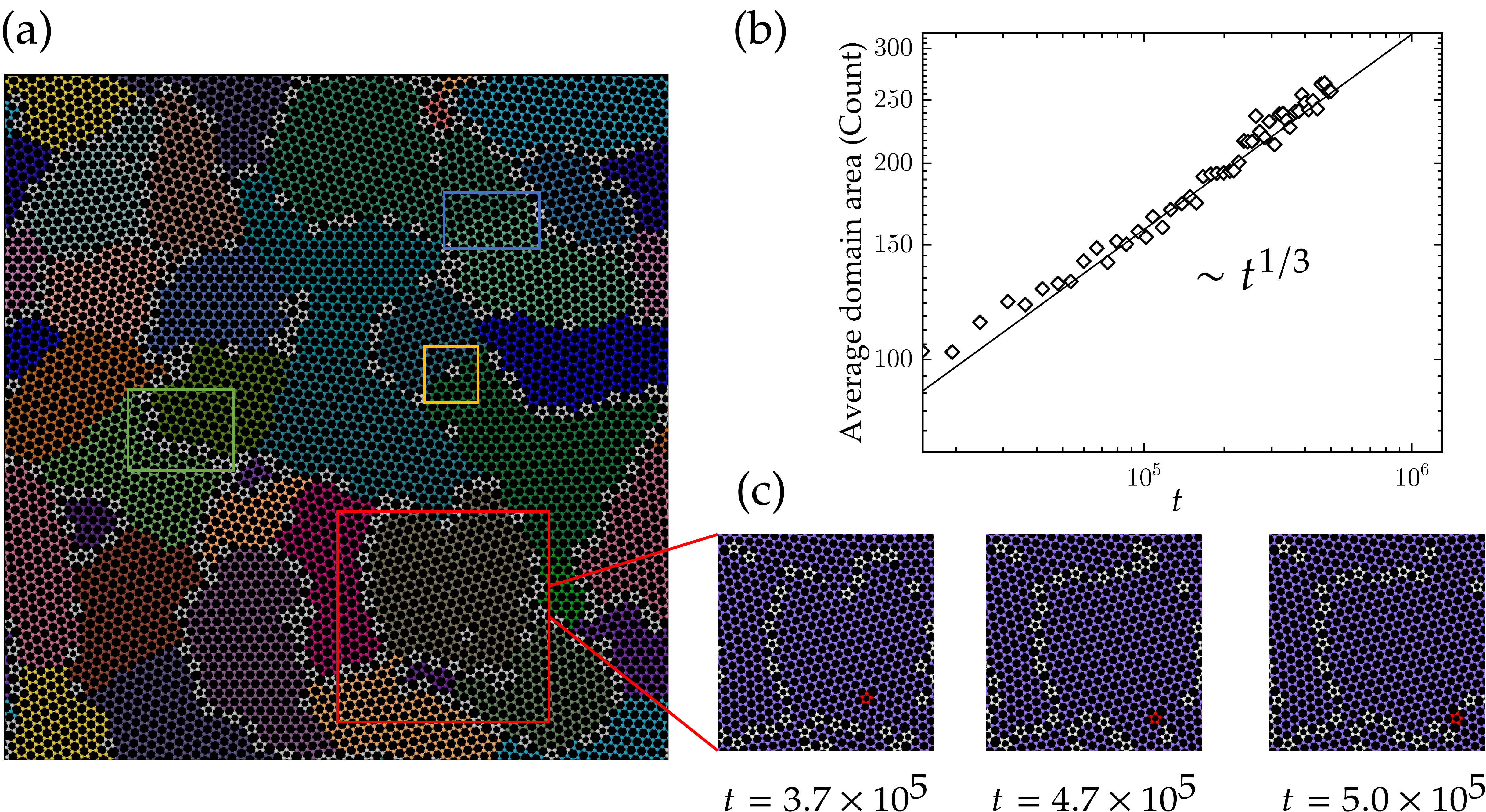}
  \caption{(a) A polycrystalline graphene with 9800 atoms, 36 crystal
    phases are identified by using the graph clustering
    algorithm.Green, yellow and blue boxes represent three different
    kind of phase separation, details see the text.(b) The average
    domain area changes in time, which approximately scales as a power
    law with $1/3$ exponent. (c) Movement of a defect island from the
    inside domain to grain boundary, showing the diffusive behavior
    of defect. \label{fig6}} 
\end{figure}
             
Our simulations show that the dynamics of the domain structure in
polycrystalline graphene are dominated by the motion of defects and
domain boundaries. As shown in Fig. \ref{fig6}(a), 36 crystal phases
(domains) were identified in a polycrystalline graphene consisting of
9800 atoms. Qualitatively, we observe various mechanisms that together
constitute the dynamics of the domain structure. First, domain
boundaries formed by strings of 5- and 7-rings separate the domains
(in the green box of Fig. \ref{fig6}(a)). These domain boundaries are
mobile, as is also observed experimentally using electron scanning
microscopy. Second, isolated defects within the crystal domain exert a
planar force on the adjacent lattices (in the yellow box of
Fig. \ref{fig6}(a)). Third, shear stress generated by grain boundaries
on both sides of the domain shears it into two fragments (in the blue
box of Fig. \ref{fig6}(a)). During the domain growth process, the
motion of defects can be classified into two scenarios. Some defects
spontaneously disappear due to energy reduction, while others undergo
diffusion motion. Figure \ref{fig6}(c) illustrates an example of defect
diffusion, where a defect island located at the center of a domain
moves to the adjacent continuous grain boundary after approximately
1.3$\times 10^5$ Monte Carlo steps.  Upon reaching the grain boundary,
it cannot cross over to the crystal domain on the other side of the
grain boundary.

We continue with a quantitative discussion of the evolution of the
domain structure. The number of atoms in the domain is used as a
representative measure of the domain area. In Fig. \ref{fig6} (b), the
average domain size is plotted as a function of time. It, square root
of average domain area, exhibits a power-law increase with an exponent
of 1/6. Given that the domains do not show a fractal structure, this
is consistent with the decay exponent of defect density shown in
Fig. \ref{fig3}(d).

At a first glance, the domain growth process in graphene resembles
that of many other systems showing Ostwald ripening.  A prototypical
domain growth process is that in the Ising model
\cite{newman1999monte}. With spin-flip (Glauber) dynamics, the
theoretical framework is known as ``Model A', in which domains of
aligned spins grow proportional to $t^{1/2}$. If the magnetization is
locally conserved, as in spin-exchange (Kawasaki) dynamics, the
theoretical framework is known as "Model B", in which these domains
grow proportional to $t^{1/3}$. In the case at hand, we do observe a
growth exponent close to 1/3, but it is less clear that a local
conservation law is active. There are a number of differences between
the domain growth in graphene and the Ising model. For instance, the
domains in graphene have a continuously varying orientation, rather
than only ``up'' and ``down'; additionally, long-ranged interactions
might play a role, especially if buckling is allowed; and while some
domain walls can easily move in some directions, the motion can be
blocked in other directions.  In future work, we hope to make a
clearer connection between domain growth in graphene and the extensive
literature on Ostwald ripening.

\subsection{Domain growth in buckled polycrystalline graphene\label{sec5}}

\begin{figure}[H]
    \centering
    \includegraphics[width=13cm]{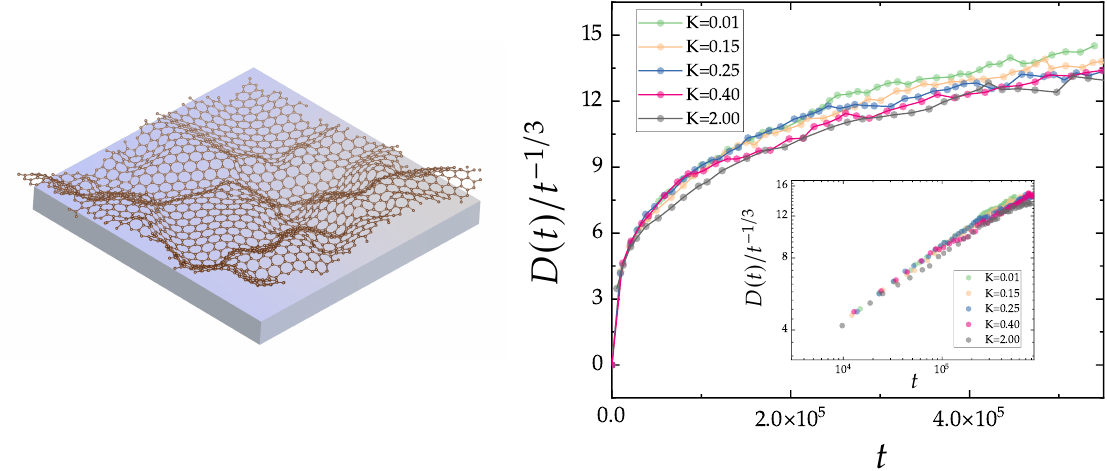}
  \caption{(a) A brief schematic diagram illustrating the growth of
    buckled polycrystalline graphene on a substrate.(b) Defect density
    of buckled polycrystalline graphene growing on various substrates
    divided by $t^{-1/3}$ (the decay rate in the flat case) over
    time.The inner figure is plotted on a double-logarithmic scale,
    demonstrating that the buckling of polycrystalline graphene slow
    down the crystallization rate.All samples are evolved start from
    an initial configuration with 20\% defects density.Since
    well-crystallized sample leads to higher buckling height $\Delta
    z$, which counteracts the suppression of substrate, the difference
    is weak for various prefactor $K$. 
    \label{fig7}}
\end{figure} 

The lowest-energy state of crystalline graphene in vacuum is a purely
two-dimensional structure.  At finite temperature already, the carbon
atoms will show out-of-plane displacements.  Once structural defects
are introduced, a free-floating layer of graphene will show even more
structure in the out-of-plane direction, a phenomenon known as
buckling. This buckling is suppressed significantly, but not
completely, if the layer of graphene is placed on a substrate.  For
the current study on domain growth, the main effect of the substrate
is the suppression of buckling.  We therefore incorporate the main
effect of the substrate by adding a harmonic confining energy term,
defined as
\begin{align}
  \label{eq6}
  {E_s} = K\sum\limits_i^N {z_i^2} 
\end{align}
Here, $N$ is the number of atoms, $z_i$ is the normal-to-plane
coordinate of the atom. The parameter $K$ sets the strength of the
interaction with substrate. Tison {\it et al.\/} \cite{tison2014grain}
have reported that the buckling resulting from defects and domain
boundaries extends to typically 5 to 20\AA; according to our previous
investigations, this corresponds to a range of $K$ values between 0.05
and 0.3. Figure \ref{fig7}(a) displays a buckled polycrystalline graphene
growing on a substrate, while Fig. \ref{fig7}(b) focuses on the
evolution of defect density in time. Specifically, the density of
non-hexagonal rings divided by $t^{-1/3}$ (the decay rate in the flat
case), for various $K$ values ranging from 0.01 to 2.00.

Notably, our findings indicate that the buckling of polycrystalline
graphene significantly slows down the domain growth process. As the
value of $K$ increases, $D(t)/t^{-1/3}$ tends to reach a constant
value in a shorter time. The difference for various $K$ is however
weak, because the higher buckling height $\Delta z$ resulting from
crystallization counteracts the suppression of substrates.  In
conclusion, flatter graphene exhibits faster coarsening.  This
intriguing observation highlights the intricate interplay between
buckling, substrate effects, and defect dynamics in the
crystallization process of graphene.

\section{Summary} \label{sec6}

In this paper, we employed a recently developed and extensively
validated model to investigate the dynamics of domain growth in
polycrystalline graphene. The dynamics consists of a sequence of
proposed bond transpositions at random locations, accepted or rejected
according to Metropolis method. The technique enables access to much
longer time scales, compared to molecular dynamics (MD) method. The
studied domain growth process is performed under zero pressure,
quenching the system from infinite temperature to approximately
3000K. The radial distribution function shows that the spatial
structures of our generated samples have good agreement with ones in
experiments at the same defect density.

Through the simulations and analysis, the dynamics revealed underlying
statistical mechanisms behind domain growth in polycrystalline
graphene.  Flat and buckled graphene are both investigated.  For the
flat case, we find that bond angles and bond lengths converged
respectively towards $120^{\circ}$ and $1.42\AA$ as a function of
time. The long-range disorder exhibited a gradual reduction, and the
defect density, represented by the proportion of non-hexagonal rings,
followed a power-law distribution with an exponent of $-1/3$ found
from our simulations over time.  In addition, the spatial correlation
of lattice orientations statistically follows a stretched exponential
form with less flat tail over times.

We identified different domains within polycrystalline graphene, and
delved into discussions regarding phase separation and defect
diffusion motion, the average domain size exhibits a power-law
increase with an exponent of $1/6$ over times. We briefly compared the
domain growth in polycrystalline graphene with the Ising dynamics. It
was found that a similar growth exponent close to 1/3 was observed in
the Kawasaki dynamics with a conserved magnetization density. However,
the domain growth in polycrystalline graphene exhibits more
complexity. Nevertheless, we believe that this correlation will
provide some guidance for our future related research.

For the buckled case, we briefly investigated the evolution of buckled
polycrystalline graphene on substrates. Our findings demonstrated
that the undulating buckling of polycrystalline graphene led to a
reduction in the crystallization rate.

Our work may provide crucial insights into the dynamics of
polycrystalline graphene during crystallization processes, which is
difficult to achieve in experiments and MD simulations. Our findings also
contribute to a deeper understanding of the development of advanced
materials and the optimization of graphene-based applications.
Moreover, the observation of reduced crystallization rates in buckled
polycrystalline graphene on substrates emphasizes the need for careful
consideration of substrate effects in future graphene-related
research.


\begin{thebibliography}{10}
	
	\bibitem{song2020tailoring}
	Ningning Song, Zan Gao, and Xiaodong Li.
	\newblock Tailoring nanocomposite interfaces with graphene to achieve high
	strength and toughness.
	\newblock {\em Science Advances}, 6(42):eaba7016, 2020.
	
	\bibitem{zhang2014fracture}
	Peng Zhang, Lulu Ma, Feifei Fan, Zhi Zeng, Cheng Peng, Phillip~E Loya, Zheng
	Liu, Yongji Gong, Jiangnan Zhang, Xingxiang Zhang, et~al.
	\newblock Fracture toughness of graphene.
	\newblock {\em Nature communications}, 5(1):3782, 2014.
	
	\bibitem{shekhawat2016toughness}
	Ashivni Shekhawat and Robert~O Ritchie.
	\newblock Toughness and strength of nanocrystalline graphene.
	\newblock {\em Nature communications}, 7(1):10546, 2016.
	
	\bibitem{balandin2008superior}
	Alexander~A Balandin, Suchismita Ghosh, Wenzhong Bao, Irene Calizo, 
	Desalegne
	Teweldebrhan, Feng Miao, and Chun~Ning Lau.
	\newblock Superior thermal conductivity of single-layer graphene.
	\newblock {\em Nano letters}, 8(3):902--907, 2008.
	
	\bibitem{chen2012thermal}
	Shanshan Chen, Qingzhi Wu, Columbia Mishra, Junyong Kang, Hengji Zhang,
	Kyeongjae Cho, Weiwei Cai, Alexander~A Balandin, and Rodney~S Ruoff.
	\newblock Thermal conductivity of isotopically modified graphene.
	\newblock {\em Nature materials}, 11(3):203--207, 2012.
	
	\bibitem{chen2008mechanically}
	Haiqun Chen, Marc~B M{\"u}ller, Kerry~J Gilmore, Gordon~G Wallace, and Dan 
	Li.
	\newblock Mechanically strong, electrically conductive, and biocompatible
	graphene paper.
	\newblock {\em Advanced Materials}, 20(18):3557--3561, 2008.
	
	\bibitem{wang2009supercapacitor}
	Yan Wang, Zhiqiang Shi, Yi~Huang, Yanfeng Ma, Chengyang Wang, Mingming Chen,
	and Yongsheng Chen.
	\newblock Supercapacitor devices based on graphene materials.
	\newblock {\em The Journal of Physical Chemistry C}, 113(30):13103--13107,
	2009.
	
	\bibitem{gwon2011flexible}
	Hyeokjo Gwon, Hyun-Suk Kim, Kye~Ung Lee, Dong-Hwa Seo, Yun~Chang Park, 
	Yun-Sung
	Lee, Byung~Tae Ahn, and Kisuk Kang.
	\newblock Flexible energy storage devices based on graphene paper.
	\newblock {\em Energy \& Environmental Science}, 4(4):1277--1283, 2011.
	
	\bibitem{chen2013terahertz}
	Pai-Yen Chen and Andrea Al{\`u}.
	\newblock Terahertz metamaterial devices based on graphene nanostructures.
	\newblock {\em IEEE Transactions on Terahertz Science and Technology},
	3(6):748--756, 2013.
	
	\bibitem{zhang2013high}
	Fan Zhang, Tengfei Zhang, Xi~Yang, Long Zhang, Kai Leng, Yi~Huang, and
	Yongsheng Chen.
	\newblock A high-performance supercapacitor-battery hybrid energy storage
	device based on graphene-enhanced electrode materials with ultrahigh energy
	density.
	\newblock {\em Energy \& Environmental Science}, 6(5):1623--1632, 2013.
	
	\bibitem{zhang2013review}
	YI~Zhang, Luyao Zhang, and Chongwu Zhou.
	\newblock Review of chemical vapor deposition of graphene and related
	applications.
	\newblock {\em Accounts of chemical research}, 46(10):2329--2339, 2013.
	
	\bibitem{mishra2016graphene}
	Neeraj Mishra, John Boeckl, Nunzio Motta, and Francesca Iacopi.
	\newblock Graphene growth on silicon carbide: A review.
	\newblock {\em physica status solidi (a)}, 213(9):2277--2289, 2016.
	
	\bibitem{xu2018liquid}
	Yanyan Xu, Huizhe Cao, Yanqin Xue, Biao Li, and Weihua Cai.
	\newblock Liquid-phase exfoliation of graphene: an overview on exfoliation
	media, techniques, and challenges.
	\newblock {\em Nanomaterials}, 8(11):942, 2018.
	
	\bibitem{jain2015strong}
	Sandeep~K Jain, Gerard~T Barkema, Normand Mousseau, Chang-Ming Fang, and
	Marijn~A van Huis.
	\newblock Strong long-range relaxations of structural defects in graphene
	simulated using a new semiempirical potential.
	\newblock {\em The Journal of Physical Chemistry C}, 119(17):9646--9655, 
	2015.
	
	\bibitem{barkema2000high}
	Gerard~T Barkema and Normand Mousseau.
	\newblock High-quality continuous random networks.
	\newblock {\em Physical Review B}, 62(8):4985, 2000.
	
	\bibitem{wooten1985computer}
	9\_F Wooten, K~Winer, and D~Weaire.
	\newblock Computer generation of structural models of amorphous si and ge.
	\newblock {\em Physical review letters}, 54(13):1392, 1985.
	
	\bibitem{d2019discontinuous}
	Federico D'Ambrosio, Vladimir Juri{\v{c}}i{\'c}, and Gerard~T Barkema.
	\newblock Discontinuous evolution of the structure of stretching
	polycrystalline graphene.
	\newblock {\em Physical Review B}, 100(16):161402, 2019.
	
	\bibitem{guenole2020assessment}
	Julien Gu{\'e}nol{\'e}, Wolfram~G N{\"o}hring, Aviral Vaid, Fr{\'e}d{\'e}ric
	Houll{\'e}, Zhuocheng Xie, Aruna Prakash, and Erik Bitzek.
	\newblock Assessment and optimization of the fast inertial relaxation engine
	(fire) for energy minimization in atomistic simulations and its
	implementation in lammps.
	\newblock {\em Computational Materials Science}, 175:109584, 2020.
	
	\bibitem{kirkwood1939skeletal}
	John~G Kirkwood.
	\newblock The skeletal modes of vibration of long chain molecules.
	\newblock {\em The Journal of Chemical Physics}, 7(7):506--509, 1939.
	
	\bibitem{liu2022structural}
	Zihua Liu, Debabrata Panja, and Gerard~T Barkema.
	\newblock Structural dynamics of polycrystalline graphene.
	\newblock {\em Physical Review E}, 105(4):044116, 2022.
	
	\bibitem{jain2015probing}
	Sandeep~K Jain, Vladimir Juricic, and Gerard~T Barkema.
	\newblock Probing crystallinity of graphene samples via the vibrational 
	density
	of states.
	\newblock {\em The journal of physical chemistry letters}, 6(19):3897--3902,
	2015.
	
	\bibitem{jain2016structure}
	Sandeep~K Jain, Vladimir Juri{\v{c}}i{\'c}, and Gerard~T Barkema.
	\newblock Structure of twisted and buckled bilayer graphene.
	\newblock {\em 2D Materials}, 4(1):015018, 2016.
	
	\bibitem{jain2017probing}
	Sandeep~K Jain, Vladimir Juri{\v{c}}i{\'c}, and Gerard~T Barkema.
	\newblock Probing the shape of a graphene nanobubble.
	\newblock {\em Physical Chemistry Chemical Physics}, 19(11):7465--7470, 
	2017.
	
	\bibitem{eder2014journey}
	Franz~R Eder, Jani Kotakoski, Ute Kaiser, and Jannik~C Meyer.
	\newblock A journey from order to disorder-atom by atom transformation from
	graphene to a 2d carbon glass.
	\newblock {\em Scientific reports}, 4(1):4060, 2014.
	
	\bibitem{larsen2016robust}
	Peter~Mahler Larsen, S{\o}ren Schmidt, and Jakob Schi{\o}tz.
	\newblock Robust structural identification via polyhedral template matching.
	\newblock {\em Modelling and Simulation in Materials Science and 
	Engineering},
	24(5):055007, 2016.
	
	\bibitem{bonald2018hierarchical}
	Thomas Bonald, Bertrand Charpentier, Alexis Galland, and Alexandre Hollocou.
	\newblock Hierarchical graph clustering using node pair sampling.
	\newblock {\em arXiv preprint arXiv:1806.01664}, 2018.
	
	\bibitem{newman1999monte}
	Mark~EJ Newman and Gerard~T Barkema.
	\newblock {\em Monte Carlo methods in statistical physics}.
	\newblock Clarendon Press, 1999.
	
	\bibitem{tison2014grain}
	Yann Tison, J{\'e}r{\^o}me Lagoute, Vincent Repain, Cyril Chacon, Yann 
	Girard,
	Fr{\'e}d{\'e}ric Joucken, Robert Sporken, Fernando Gargiulo, Oleg~V Yazyev,
	and Sylvie Rousset.
	\newblock Grain boundaries in graphene on sic (0001) substrate.
	\newblock {\em Nano letters}, 14(11):6382--6386, 2014.
	
\end{thebibliography}
\end{document}